\documentclass[12pt]{article}
\usepackage{graphics,epsfig}
\usepackage{graphicx}

\usepackage{stmaryrd}
\usepackage{mathrsfs}
\usepackage{amssymb}
\normalsize
\newcommand{\bea}{\begin{eqnarray}}
\newcommand{\ena}{\end{eqnarray}}
\newcommand{\vs}[1]{\vspace{#1 mm}}

\begin{document}
\addtolength{\baselineskip}{.35mm}
\newlength{\extraspace}
\setlength{\extraspace}{2.5mm}
\newlength{\extraspaces}
\setlength{\extraspaces}{2.5mm}
\begin{titlepage}
\begin{center}

\vspace{3.5cm}

\begin{flushright}
KU-TP 026 \\
\end{flushright}
\vs{5}

{\Large \bf{Shear Viscosity from Gauss-Bonnet Gravity with a Dilaton Coupling}}\\[1.5cm]

{Rong-Gen Cai $^{a,}$\footnote{Email: cairg@itp.ac.cn},}
{Zhang-Yu Nie $^{a,b,}$\footnote{Email: niezy@itp.ac.cn},}
Nobuyoshi Ohta $^{c,}$\footnote{Email: ohtan@phys.kindai.ac.jp} and
{Ya-Wen Sun $^{a,b,}$\footnote{Email: sunyw@itp.ac.cn},}

\vspace*{0.5cm}

{\it $^{a}$ Key Laboratory of Frontiers in Theoretical Physics,
Institute of Theoretical Physics, Chinese Academy of Sciences,
P.O.Box 2735, Beijing 100190, China

$^{b}$ Graduate University of Chinese Academy of Sciences,
\\ YuQuan
Road 19A, Beijing 100049, China

$^c$ Department of Physics, Kinki University, Higashi-Osaka, Osaka
577-8502, Japan }
\date{\today}
\vspace{2.5cm}

\textbf{Abstract} \vspace{5mm}

\end{center}

We calculate the shear viscosity of field theories with gravity
duals of Gauss-Bonnet gravity with a non-trivial dilaton using
AdS/CFT.  We find that the dilaton filed has a non-trivial
contribution to the ratio of shear viscosity over entropy density
and after imposing causal constraint for the boundary field theory,
the new lower bound $4/25\pi$, obtained from pure Gauss-Bonnet
gravity,  may have a small violation.

\end{titlepage}
\newpage
\section{Introduction}
The development of AdS/CFT correspondence
\cite{Maldacena:1997re,{Gubser:1998bc},{Witten:1998qj},{Aharony:1999ti}}
provides an efficient way to study the hydrodynamic properties of
strongly coupled gauge field theories. A remarkable example is the
calculation of the shear viscosity
\cite{Policastro:2001yc,{Kovtun:2003wp},{Buchel:2003tz},{Kovtun:2004de}}
for conformal field theories with gravity dual descriptions. The
ratio of the shear viscosity over entropy density is calculated to
be $1/4\pi$ for a large variety of conformal field theories with
gravity duals in the large N limit, with or without chemical
potentials
\cite{{Mas:2006dy},{Son:2006em},{Saremi:2006ep},{Maeda:2006by},{Cai:2008in}}.
With large N corrections, the ratio of shear viscosity over entropy
density was found to have a positive correction to $1/4\pi$
\cite{Buchel:2004di,{Benincasa:2005qc},{Buchel:2008ac},{Buchel:2008wy},{Buchel:2008sh},{Myers:2008yi},{Buchel:2008ae}}.
Along with the fact that all known substances including water and
liquid helium as well as the quark-gluon plasma created at
Relativistic Heavy Ion Collider (RHIC) have a larger shear viscosity
over entropy density ratio than $1/4\pi$, it was conjectured that
$1/4\pi$ is a universal lower bound for all materials, which is
called the Kovtun-Starinets-Son (KSS) bound
\cite{{Policastro:2002se},{Policastro:2002tn},{Buchel:2004qq},{Cohen:2007qr},{Son:2007vk},{Cherman:2007fj},
{Chen:2007jq},{Son:2007xw},{Fouxon:2008pz},{Dobado:2008ri},{Landsteiner:2007bd}}.

However, in Refs.~\cite{Brigante:2007nu,{Brigante:2008gz},{KP}} the
authors considered $R^2$ higher derivative gravity corrections and
found that the modification of the ratio of shear viscosity over
entropy density to the conjectured bound is negative, which means
that the lower bound is violated in this condition. The higher
derivative gravity corrections they considered can be seen as
generated from stringy corrections given the vastness of the string
landscape. A new lower bound, $4/25\pi$ which is smaller than
$1/4\pi$, is proposed, based on the causality of dual field theory.

In Refs.~\cite{Brustein:2008cg,{Brustein:2008xx}} it was conjectured
that the shear viscosity is fully determined by the effective
coupling of the transverse gravitons on the horizon in the dual gravity
description. This was confirmed in Ref.~\cite{Iqbal:2008by} using
the scalar membrane paradigm approach and also in
Ref.~\cite{Cai:2008ph} by directly calculating the on-shell action
of the transverse gravitons. In these calculations, the effective action
of the transverse gravitons in a given background in generalized
gravity theories was assumed to be a minimally coupled massless
scalar with an effective coupling which depends on the radial coordinate,
while in the Einstein gravity, the effective coupling is just a constant.
However, it cannot be directly seen just from the action of the dual
gravity theory whether the effective action of the transverse gravitons
really takes the form assumed there. Furthermore the choice of coordinate
system of the background black hole geometry also affects the form
of the action of the transverse gravitons because this formalism is
not covariant under coordinate transformations. Thus to consider
more general gravity theories, we use an effective action of
the transverse gravitons with a slightly generalized form of coupling.
In this new formalism, we define a new three-dimensional effective
metric $\widetilde{g}_{\mu\nu}$ and the transverse gravitons are
minimally coupled to this new effective metric. Following the same
procedure in Ref.~\cite{Cai:2008ph}, we find that the shear
viscosity of the dual field theory can also be calculated and it is
no longer fully determined by the effective coupling of gravitons in
the general case.

The concrete expression of the action of the transverse
gravitons for a given gravity theory still needs to be calculated
explicitly. For the Einstein and Gauss-Bonnet gravity with scalars and
vectors coupled only to the ordinary derivatives of metrics,
the expression of the effective action of the transverse gravitons has been
found to have the same dependence on the background metric as the pure
Einstein and Gauss-Bonnet gravity, respectively~\cite{Cai:2008ph}.
 However, this class of
the theories are not the low-energy effective theories of the strings,
which always contain the dilaton with nontrivial couplings.
In this paper, we calculate the effective action of the transverse gravitons
for Gauss-Bonnet gravity coupled to a nontrivial dilaton
field~\cite{Guo:2008eq}, and examine how the above results are modified.
In fact we will find that the effective action of the transverse gravitons
is not the same as pure Gauss-Bonnet gravity.

In this nontrivial example, after careful analysis we find that the
causal constraint that should be imposed to make sure the boundary
field theory does not violate causality is still simple. With the
constraint, we find that the ratio of $\eta/s$ may have a small
violation to the new lower bound proposed in
Ref.~\cite{Brigante:2007nu}.

In the remainder of this paper, we first give a brief
calculation of the shear viscosity given the form of the effective
action of the transverse gravitons in Sec. 2. In Sec. 3 we calculate the
effective metric for Gauss-Bonnet gravity coupled with a nontrivial
dilaton field. In Sec. 4 we give the causal constraints and the
analysis of the value of the ratio.  Sec. 5 gives the conclusions
and discussions.

\section{Shear viscosity}

In this section we calculate the shear viscosity of the dual field
theory given the form of the effective action of the transverse
gravitons. We use the Kubo formula
\begin{equation}\label{eta}
\eta=\lim_{\omega\rightarrow 0}\frac{1}{2\omega
i}\Big(G^A_{xy,xy}(\omega,0)-G^R_{xy,xy}(\omega,0)\Big),
\end{equation}
where $\eta$ is the shear viscosity and the retarded Green's
function is defined by
\begin{equation}
G^R_{\mu\nu,\lambda\rho}(k)=-i\int d^4xe^{-ik\cdot x}\theta (t)
\langle[T_{\mu\nu}(x),T_{\lambda\rho}(0)] \rangle.
\end{equation}
These are defined on the field theory side. The advanced Green's
function can be related to the retarded Green's function of energy
momentum tensor by
$G^A_{\mu\nu,\lambda\rho}(k)=G^R_{\mu\nu,\lambda\rho}(k)^{*}$. Using
the field operator correspondence, the Green function of energy
momentum tensors of the field theory can be calculated through the
effective action of gravitons of the dual gravity theory.

For simplicity, here we choose spatial coordinates so that the
momentum of the perturbation points along the $z$-axis. The
perturbations can be written as $h_{\mu\nu}=h_{\mu\nu}(t,z,u)$ with the radial
coordinate $u$. In
this basis, there are three groups of gravity perturbations, each of
which is decoupled from the others: the scalar, vector and tensor
perturbations~\cite{Kovtun:2005ev}. Here we use the simplest one,
the tensor perturbation $h_{xy}$. We use $\phi$ to denote this
perturbation with one index raised $\phi=h^x_y$ and write $\phi$ in
a basis as $\phi(t,u,z)=\phi(u)e^{-i\omega t+ip z}$.

We can then get the effective action of this transverse graviton by
keeping terms to the second order of $\phi$ in the gravity action.
For the Einstein gravity, the effective action of the transverse gravitons
is always
\begin{equation}\label{b}
S=\frac{1}{16\pi G}\bigg(-\frac{1}{2}\bigg)\int d^5x\sqrt{-g}
(g^{\mu\nu}\nabla_{\mu} \phi\nabla_{\nu}\phi),
\end{equation}
when matter fields are coupled to the metric minimally
\cite{Cai:2008ph}, where $g_{\mu\nu}$ is the metric of the
background black hole solution. For general gravity theories coupled
to matter fields minimally or non-minimally, the action of
the transverse gravitons is no longer of the form (\ref{b}). In
Refs.~\cite{Iqbal:2008by,{Cai:2008ph}} a deformed form of the effective
action
\begin{equation}\label{c}
S=\frac{1}{16\pi G}\int d^5x\sqrt{-g}K_{eff}(u)
(g^{\mu\nu}\nabla_{\mu} \phi\nabla_{\nu}\phi),
\end{equation}
is studied, where $K_{eff}(u)$ is an effective coupling constant.
There might be some extra terms appearing in (\ref{c}) like
$N(u)\partial_{z}\phi\partial_{z}\phi$ where $N(u)$ is a function
regular at $u=1$ (black hole horizon), and these terms will not
affect the value of $\eta$. With the addition of the extra terms,
the $K_{eff}(u)$ in (\ref{c}) is not a scalar under the general
coordinate transformation, so the expression in (\ref{c}) is not a
real covariant form. Generally, the factors in front of
$g^{tt}\partial_{t}\phi\partial_{t}\phi$ and
$g^{uu}\partial_{u}\phi\partial_{u}\phi$ are not always the same and
in these cases the effective action of the transverse gravitons cannot
be written in the form of (\ref{c}). It is difficult to determine
whether the effective action of the transverse gravitons takes
such a form as in (\ref{c}) for a generic gravity theory. Thus in
the following of this section, we use a general form of the expression
of the effective action of the transverse gravitons, which is valid
generally and does not depend on the choice of coordinate system.

For gravity theories in which the transverse gravitons can be
decoupled from other perturbations, a general form of the effective
bulk action of the transverse gravitons can be written as
\begin{equation}\label{usual}
S=\frac{V_{x,y}}{16\pi G}\bigg(-\frac{1}{2}\bigg)\int
d^3x\sqrt{-\widetilde{g}} (\widetilde{K}(u){\widetilde{g}}^{MN}
\widetilde{\nabla}_{M}\phi\widetilde{\nabla}_{N}\phi+m^2\phi^2)
\end{equation}
up to some total derivatives, where $\widetilde{g}_{MN}$ is a
three-dimensional effective metric, $m$ is an effective mass which
can be any function of the radial coordinate and
$\widetilde{\nabla}_{M}$ is the covariant differential using the
metric $\widetilde{g}_{MN}$. Here we use the effective metric
$\widetilde{g}_{MN}$ to denote the factor in front of
$\widetilde{\nabla}_{M}\phi\widetilde{\nabla}_{N}\phi$ in the action
of graviton.  $\phi=h^x_y$ is a scalar in the three dimensions of
$t$, $u$ and $z$, while it is not a scalar in the whole
five-dimensional system. Because we have assumed that $h^x_y$ only
depends on the coordinates $t$, $u$ and $z$, the effective action of
$h^x_y$ can be viewed as a deduced three-dimensional action where
the other two directions can be integrated out. Thus this is not the
dimensional reduction in the usual sense, and the Newton constant in
(\ref{usual}) is still the five-dimensional Newton constant. We
write the action in the three-dimensional form so that the action
(\ref{usual}) takes a general covariant form and $\widetilde{K}(u)$
is a scalar under general coordinate transformations. In the
following of this paper, we use $g_{\mu\nu}$, ($\mu$, $\nu=t$, $u$,
$x$, $y$, $z$) to denote the metric of the five-dimensional
background and $\widetilde{g}_{MN}$, ($M$, $N=t$, $u$, $z$) to
denote the three-dimensional effective metric. The effective action
for the transverse gravitons in the Einstein gravity can be obtained
from (\ref{usual}) by choosing $\widetilde{g}^{MN}=g^{MN}$ for $M$,
$N=t$, $u$, $z$ and $K(u)=\sqrt{-g}/\sqrt{-\widetilde{g}}$. Thus in
this case $\widetilde{K}(u)$ comes from the determinant of the
metric of the $x$ and $y$ directions. Note that here
$\widetilde{K}(u)$ is not the same one as $K_{eff}(u)$ in (\ref{c}).
In fact this $\widetilde{K}(u)$ can be absorbed into
$\widetilde{g}_{MN}$ and be eliminated to give a minimally coupled
action
\begin{equation}\label{minimal}
S=\frac{V_{x,y}}{16\pi G}\bigg(-\frac{1}{2}\bigg)\int
d^3x\sqrt{-\overline{{g}}} ({\overline{{g}}}^{MN}\partial_{M}
\phi\partial_{N}\phi+\bar{m}^2\phi^2),
\end{equation}
with
\begin{equation}\label{gbar}
\overline{g}^{MN}=\widetilde{K}^{-2}(u)\widetilde{g}^{MN},
\end{equation}
and
\begin{equation}\label{mbar}
\bar{m}^2=\widetilde{K}(u)^{-3}m^2.
\end{equation}
 In the following
calculation of this and the next section, we will still keep
$\widetilde{K}(u)$ in order to have $\widetilde{g}^{MN}=g^{MN}$ for
$M$, $N=t$, $u$, $z$ in the case of the Einstein gravity.

Before we continue to calculate the shear viscosity, we first write
down the background metric. In this paper we mainly focus on the
case of Ricci-flat black hole backgrounds. The case for black holes
with hyperbolic horizon topology has been discussed
recently~\cite{{Neupane:2008dc},{Koutsoumbas:2008wy}}. We assume
that the background black hole solution is of the form
\begin{equation}\label{black}
ds^2=-{g(u)(1-u)}dt^2+\frac{1}{h(u)(1-u)}du^2+\frac{r_+^2}{ul^2}(d\vec{x}^2),
\end{equation}
where the horizon of the black hole locates at $u=1$ and the AdS
boundary is at $u=0$, $h(u)$, $g(u)$ are functions of $u$, regular
at $u=1$ and $l$ is related to the cosmological constant $\Lambda$
by $l=\sqrt{-6/\Lambda}$. $r_+$ is the black hole horizon radius in
the radial coordinate $r$, which has a relation to the coordinate
$u$ used here through $u=\frac{r_+^2}{r^2}$. In the usual Einstein
gravity with the cosmological constant $\Lambda$, $l$ is just the
AdS radius, but it could be different from the effective AdS radius
in more general gravity theories, for example, Einstein-Gauss-Bonnet
theory with a dilaton~\cite{Guo:2008eq}, which we will consider
below. Of course, one may take $l$ as the effective AdS radius in
those gravity theories, and this will not change the result. In what
follows, for simplicity, we will keep $l=\sqrt{-6/\Lambda}$ even in
the case of Einstein-Gauss-Bonnet theory with a dilaton. In
addition,  let us mention  that $h(u)$ and $g(u)$ should be regular
at the horizon. This implies that the Ricci-flat black hole solution
we consider here is a nonextremal one.

Here we choose a diagonal background metric and we will argue that
generally the effective three dimensional metric
$\widetilde{g}_{MN}$ which appears in the effective action of
transverse gravitons should also be diagonal even when Gauss-Bonnet
terms are present. Gauss-Bonnet gravity as well as general Lovelock
gravity have the property that the Einstein equations of motion
contain at most second derivatives of the metric. Thus in the
effective action of transverse gravitons (\ref{minimal}) for pure
Gauss-Bonnet gravity there are also at most second derivatives of
$\phi$, where higher derivative terms can only exist in the total
derivative terms which do not affect our result. To get the
effective action of transverse gravitons, we should expand the
gravity action to the second order of $\phi$ and with the reason
stated above we only need to keep terms up to the second derivatives
of $\phi$. By detailed calculation of the expansion of Riemann
tensors with respect to $\phi$, we can see that the possible
non-zero coefficients of non-diagonal terms can only come from the
Gauss-Bonnet term in the case of Gauss-Bonnet gravity. We give a
simple illustration here to show that these coefficients are all
zero given the assumption that the background metric is diagonal. We
denote the three components of the Gauss-Bonnet term by
$R_{a_1a_2a_3a_4}R_{a_5a_6a_7a_8}g^{b_1b_2}g^{b_3b_4}g^{b_5b_6}g^{b_7b_8}$,
where the indices $a_i$ and $b_i=t,$ $u,$ $x$, $y$, $z$ should be
contracted for $i=1,\cdots,8$ and the metric here is the background
metric. There are three kinds of perturbations of these terms which
can give non-diagonal terms in (\ref{minimal}) and these three kinds
of perturbations can be written out as
$\delta_{(1)}R_{a_1a_2a_3a_4}\delta_{(1)}R_{a_5a_6a_7a_8}g^{b_1b_2}g^{b_3b_4}g^{b_5b_6}g^{b_7b_8}$,
$\delta_{(2)}R_{a_1a_2a_3a_4}R_{a_5a_6a_7a_8}g^{b_1b_2}g^{b_3b_4}g^{b_5b_6}g^{b_7b_8}$,
and
$\delta_{(1)}R_{a_1a_2a_3a_4}R_{a_5a_6a_7a_8}\delta_{(1)}g^{b_1b_2}g^{b_3b_4}g^{b_5b_6}g^{b_7b_8}$,
where $\delta_{j}$ means to keep terms of the $j$-th order of
$\phi$. By careful analysis of the non-zero components of
$\delta_{(j)}R_{a_1a_2a_3a_4}$, we find that for the case of the
first two kinds of perturbations, there are always indices which
appear an odd number of times in $a_i$ while because the background
metric is diagonal, all indices should appear an even number of
times in $b_i$ and for the last kind of perturbation, some indices
appear an odd number of times in $a_i$ and $b_1$, $b_2$ while they
have to appear an even number of times in $b_i$ for $i$ from $3$ to
$8$. Thus the indices can't be properly contracted and all these
three kinds of perturbations can't exist, so the effective metric of
$\widetilde{g}_{MN}$ should be diagonal. When matter fields are
coupled to the Gauss-Bonnet term, as long as the Einstein equation
of motion contains at most second derivatives of the metric
component $g_{xy}$ as a function of $t$, $u$, $z$ and the
perturbation of the transverse gravitons can get decoupled from the
perturbation of matter fields, the arguments above are still valid.
Here the Einstein equation of motion can contain higher derivative
terms by counting both the matter fields and the metric but it can
only contain at most second derivative terms of the metric component
$g_{xy}$. In the following of this paper, we only consider this kind
of gravity theory. This argument can also be used to show that
$\widetilde{g}_{MN}$ would also be diagonal for Lovelock gravity,
which we do not display explicitly here.

We then follow the procedure in Ref.~\cite{Cai:2008ph} to calculate
$\eta$. We write the action of the transverse gravitons in the momentum
space
\begin{equation}\label{action1}
S=\frac{V_{x,y}}{16\pi G}\bigg(-\frac{1}{2}\bigg)\int
\frac{dwdp}{(2\pi)^2} du\sqrt{-\widetilde{g}}
\bigg(\widetilde{K}(u)(\widetilde{g}^{uu}\phi'\phi'
+w^2\widetilde{g}^{tt}\phi^2+p^2\widetilde{g}^{zz}\phi^2)+m^2\phi^2\bigg),
\end{equation}
by expanding
\begin{equation}
\phi(t,u,z)=\int
\frac{dwdp}{(2\pi)^2}\phi(u;k)e^{-iwt+ipz},~~~k=(w,0,0,p),~~~\phi(u;-k)=\phi^*(u;k),
\end{equation}
where the prime stands for the derivative with respect to $u$ and
the $\phi^2$ terms should be recognized as $\phi^*\phi$. For the
Einstein gravity, $\widetilde{g}_{MN}=g_{MN}$ for $M$, $N=t$, $u$,
$z$, $\widetilde{K}(u)=r_+^2/ul^2$ and $m=0$. Generally,
$\widetilde{K}(u)$ is a regular function at $u=1$ and
$\widetilde{g}_{MN}$ is a diagonal metric similar to the background
metric. The equation of motion of the transverse gravitons can be
obtained from the action (\ref{action1}) as
\begin{equation}\label{a}
\phi''(u,k)+A(u)\phi'(u,k)+B(u)\phi(u,k)=0,
\end{equation}
where
\begin{equation}
A(u)=\frac{(\sqrt{-\widetilde{g}}\widetilde{K}(u)\widetilde{g}^{uu})'}
{\sqrt{-\widetilde{g}}\widetilde{K}(u)\widetilde{g}^{uu}},
\end{equation}
and
\begin{equation}
B(u)=-\widetilde{g}_{uu} \Big( \widetilde{g}^{tt}w^2
+ \widetilde{g}^{zz}p^2+\frac{m^2}{\widetilde{K}(u)} \Big).
\end{equation}
In order to solve the equation, we should make a detailed analysis
of the property of $\widetilde{g}_{uu}$, $\widetilde{g}_{tt}$, and
$B(u)$. The gravity action we consider should at least contain an
Einstein term $\sqrt{-g}R$ and there may be other terms like
$\sqrt{-g}R_{GB}^2$ and so on. The contribution to the effective
action of the transverse gravitons can be obtained separately from these
action terms and they should be summed. Thus $\widetilde{g}_{MN}$ should
at least contain a contribution of $g_{MN}$ which comes from the Einstein term.
Generally $\widetilde{g}_{uu}$ and $\widetilde{g}^{tt}$ should have
poles at most of the same order as $g_{uu}$ and $g^{tt}$, i.e. poles of
the first order. We isolate the parts of the first order pole in
$\widetilde{g}_{uu}$ and $\widetilde{g}^{tt}$ denoted as
$\widetilde{g}_{uu}^1/(1-u)$ and $\widetilde{g}^{tt}_1/(1-u)$ respectively,
where $\widetilde{g}_{uu}^1$ and $\widetilde{g}^{tt}_1$ are finite at $u=1$.
We then follow the standard procedure to solve Eq.~(\ref{a}).
First we impose the incoming boundary condition at the horizon so that
\begin{equation}
\phi(u)=(1-u)^{-i\beta w}F(u),
\end{equation}
where $F(u)$ is regular at the horizon. $\beta$ can be calculated by
considering (\ref{a}) in the limit $u\rightarrow1$, which leads to
\begin{equation}\label{beta}
\beta=\sqrt{-\widetilde{g}_{uu}^1\widetilde{g}^{tt}_1}\Big|_{u=1}.
\end{equation}
Note here that $\widetilde{g}^{zz}$ does not have any poles just
like $g^{zz}$, so it does not affect the value of $\beta$. Also we
assume $m^2$ has poles at most of the first order at the horizon, so
$m^2$ does not affect the value of $\beta$ either. Because we only
need to know the $w\rightarrow0$ behavior of this graviton, we can
expand the solution as
\begin{equation}
F(u)=1+i\beta wF_0(u)+O(w^2)+O(p^2).
\end{equation}
By expanding Eq.~(\ref{a}) to the first order of $w$, we
get the equation of $F_{0}(u)$:
\begin{equation}
F_0''(u)+A(u)F_0'(u)+\frac{1}{(1-u)^2}+\frac{A(u)}{1-u}=0.
\end{equation}
The solution of this function is already given in Ref.~\cite{Cai:2008ph}
and it can be uniquely determined by imposing the regular boundary conditions.
We can determine the derivative of the solution as
\begin{equation}
F_{0}'(u)=\frac{S(1)}{1-u}\bigg(\frac{1}{S(u)}-\frac{1}{S(1)}\bigg),
\end{equation}
where
$S(u)=(\sqrt{-\widetilde{g}}\widetilde{K}(u)\widetilde{g}^{uu})/(1-u)$
and $S(1)=\lim_{u\rightarrow 1}S(u)$. Using the same arguments
as in the appendix of Ref.~\cite{Cai:2008ph}, we find that the on-shell action is
\begin{equation}
S_{on-shell}=\frac{V_{x,y}}{16\pi G}\bigg(-\frac{1}{2}\bigg)\int
\frac{dwdp}{(2\pi)^2} du
\bigg(\sqrt{-\widetilde{g}}\widetilde{K}(u)\widetilde{g}^{uu}\phi'\phi\bigg)'
\end{equation}
after the Gibbons-Hawking surface contribution has been counted.
Integrating this action gives
\begin{equation}
S_{on-shell}=\frac{V_{x,y}}{16\pi G}\bigg(-\frac{1}{2}\bigg)\int
\frac{dwdp}{(2\pi)^2}
\bigg(\sqrt{-\widetilde{g}}\widetilde{K}(u)\widetilde{g}^{uu}\phi'\phi\bigg)
\bigg|_{u=1}^{u=0}.
\end{equation}
Then the shear viscosity can be calculated using the Kubo formula
(\ref{eta}) to be
\begin{equation}
\eta=\frac{1}{16\pi G}\lim_{w\rightarrow
0}\frac{-\sqrt{-\widetilde{g}}\widetilde{K}(u)\widetilde{g}^{uu}\phi'\phi|_{u=0}}{iw}=\frac{1}{16\pi
G}(\beta S(1)).
\end{equation}
After substituting $\beta$ in (\ref{beta}) and $S(u)$ we find that
\begin{equation}\label{etanew}
\eta=\frac{1}{16\pi
G}\Big(\sqrt{\widetilde{g}_{zz}}\widetilde{K}(u)\Big)\Big|_{u=1}.
\end{equation}
For the black hole backgrounds where the area formula of black hole
entropy still holds, the entropy density is
\begin{equation}
s=\frac{1}{4G}\frac{r_+^3}{l^3},
\end{equation}
and the ratio of shear viscosity over entropy density is
\begin{equation}
\label{e23}
\frac{\eta}{s}=\frac{1}{4\pi}\frac{l^3}{r_+^3}\Big(\sqrt{\widetilde{g}_{zz}}
\widetilde{K}(u)\Big)\Big|_{u=1}.
\end{equation}
Thus we obtained a general formula to calculate the ratio of shear
viscosity over entropy density. It can be checked that for those
actions which can be written in the form (\ref{c}), the ratio of
$\eta/s$ reproduces the dependence on the effective coupling on the
horizon obtained in Ref.~\cite{Cai:2008ph}.

\section{Non-trivial dilaton}

In the previous section, we have given a general formula for the
shear viscosity of the dual field theory. Using this formula, we can
read the ratio of shear viscosity over entropy density from the
effective action of the transverse gravitons in the gravity description.
However, the exact form of the effective action of the transverse
gravitons still has to be calculated case by case. In
Ref.~\cite{Cai:2008ph} we know that for the Einstein and Gauss-Bonnet
gravity coupled with matter fields minimally, the effective action
of the transverse gravitons is not affected by the matter fields, but
the arguments in Ref.~\cite{Cai:2008ph} do not apply to gravity
theories with non-minimally coupled matter fields. In this section
we calculate the effective action of the transverse gravitons for
Gauss-Bonnet gravity with a non-minimally coupled dilaton and use
the formula (\ref{etanew}) to obtain the shear viscosity of the dual
field theory.

The action we consider in this section is \cite{Guo:2008eq}
\begin{equation}\label{dilaton}
S=\frac{1}{16\pi G}\int d^5x
\sqrt{-g}\bigg[R-\frac{1}{2}\nabla_{\mu}\phi_d\nabla^{\mu}\phi_d+\frac{\lambda
l^2}{2}e^{-\gamma\phi_d}(R_{\mu\nu\rho\sigma}R^{\mu\nu\rho\sigma}
-4R_{\mu\nu}R^{\mu\nu}+R^2)-2\Lambda e^{\tau\phi_d}\bigg],
\end{equation}
where $\lambda$ is the Gauss-Bonnet coupling, $l$ is the AdS radius,
$\gamma$ and $\tau$ are constants and $\phi_d$ is the
dilaton field. This action can be obtained by transforming the
string frame action with a nontrivial dilaton to the Einstein
frame. The ten-dimensional critical string theory predicts
$\gamma=1/2$ and in this paper we follow \cite{Guo:2008eq} to leave
$\gamma$ unfixed. The term $2\Lambda e^{\tau\phi_d}$ is the
effective cosmological term produced by a nontrivial
dilaton, and $\tau=5/2$ if we assume that the term comes from the
expectation value of the RR 10-form in type IIB superstring theory.
However there are also other sources of this term and here we leave
$\tau$ unfixed also. In fact the effective cosmological
term can be replaced by a more general potential $V(\phi)$,
which does not affect the result. The unique requirement is to have
an asymptotically AdS black hole solution with such a potential.
% For this specific choice of the dilaton potential, we can define an
% effective AdS radius $\widetilde{l}$ to be
% \begin{equation}
% \widetilde{l}=e^{-\frac{1}{2}\tau\phi_d}l,
% \end{equation}
% which changes along the radial coordinate $u$.
 In addition, to make sure that the
gravity regime is valid, we have to impose the condition that
$\lambda \ll 1$, $\phi_d$ not too large and $\phi_d$ changes slowly
along the radial coordinate $u$.

We also assume the Ricci-flat black hole solution (\ref{black}) and
calculate the effective action of the transverse gravitons on this
background. It can be checked directly from the first-order Einstein
and Klein-Gordon equation of motion that the transverse gravitons
can get decoupled from other perturbations. Then we can get the
effective action of transverse gravitons by keeping quadratic terms
of $\phi$ in the original action (\ref{dilaton}) and find that we
have at most second derivatives of $\phi$ up to this order. Thus the
action of the transverse gravitons can be written in the form
(\ref{usual}) and the effective three-dimensional metric for this
specific case is
\begin{equation}\label{uu}
\widetilde{g}^{uu}=\bigg[1+ \frac{\lambda
l^2}{2}e^{-\gamma\phi_d}\bigg(\frac{g^{uu}g_{tt}'}{ug_{tt}}+\frac{2\gamma\phi_d'
 g^{uu}g_{tt}'}{g_{tt}}-\frac{2\gamma
 g^{uu}\phi_d'}{u}\bigg)\bigg]g^{uu},
\end{equation}
\begin{equation}\label{tt}
\widetilde{g}^{tt}=\bigg[1+\frac{\lambda l^2}{2} e^{-\gamma \phi_d}
\bigg( \frac{{g^{uu}}'}{u} -\frac{3g^{uu}}{u^2} -\frac{2\gamma
g^{uu}\phi_d'}{u} -4\gamma ^2 g^{uu} {\phi_d'}^2+4\gamma
g^{uu}\phi_d''+2\gamma\phi_d'{g^{uu}}'\bigg) \bigg]g^{tt},
\end{equation}
and
\begin{eqnarray}\label{zz}
\widetilde{g}^{zz}=\bigg[1+\frac{\lambda
l^2}{2}e^{-\gamma\phi_d}\bigg(-\frac{{g^{uu}} '{g_{tt}}
'}{g_{tt}}+\frac{g^{uu}{{g_{tt}'}^2}}{g_{tt}^2} +\frac{2\gamma
g^{uu} {g_{tt}}'\phi_d'}{g_{tt}}-\frac{2g^{uu}{g_{tt}}''}{g_{tt}}
+4\gamma g^{uu} {\phi_d}''\nonumber
\\+2\gamma {g^{uu}}'{\phi_d}'-4\gamma
^2g^{uu}{\phi_d'}^2 \bigg)\bigg]g^{zz},
\end{eqnarray}
where the metric $g^{\mu\nu}$ denotes the background metric. After a
lengthy calculation, it is found that the $m^2$ term vanishes if we
use the Einstein equations for the background metric.

In fact, the effective action of the transverse gravitons for this
gravity theory (\ref{dilaton}) can  be written as
%\begin{equation}\end{equation}
\begin{equation}
S=\frac{1}{16\pi G}\bigg(-\frac{1}{2}\bigg)\int d^5 x
\sqrt{-g}\widetilde{g}^{\mu\nu}\partial_{\mu}\phi\partial_{\nu}\phi,
\end{equation}
where $g^{\mu\nu}$ here is a five dimensional metric with
$\widetilde{g}^{\mu\nu}=\widetilde{g}^{\mu\nu}$ for $\mu$, $\nu=t$,
$u$, $z$ and $\widetilde{g}^{\mu\nu}=g^{\mu\nu}$ for $\mu$, $\nu=x$,
$y$, as a non-covariant form, so
$\widetilde{K}(u)=\sqrt{-g}/\sqrt{-\widetilde{g}}$, where
$g^{\mu\nu}$ denotes the background five-dimensional metric
(\ref{black}). Thus the shear viscosity can be obtained using the
formula (\ref{etanew}) as
\begin{equation}
\eta=\frac{1}{16\pi G}\frac{r_+^3}{l^3}\bigg(1-\frac{\lambda
l^2}{2}e^{-\gamma\phi_d(1)}h(1)\Big(1+2\gamma{\phi_d}'(1)\Big)\bigg).
\end{equation}
For this Ricci-flat black hole, the entropy still obeys the
Bekenstein-Hawking entropy area law \cite{Cai,{Guo:2008eq}}, so the
entropy density can be easily written as
%\begin{equation}\end{equation}
\begin{equation}
s=\frac{1}{4G}\frac{r_+^3}{l^3}.
\end{equation}
Thus the ratio of shear viscosity over entropy density is
\begin{equation}\label{etadi}
\frac{\eta}{s}=\frac{1}{4\pi}\bigg(1-\frac{\lambda
l^2}{2}e^{-\gamma\phi_d(1)}h(1)\Big(1+2\gamma{\phi_d}'(1)\Big)\bigg).
\end{equation}
Here the dilaton $\phi_d$ should be regular at the horizon. The
formula (\ref{etadi}) for $\eta/s$ is valid even when other scalar
or vector fields are present as long as those fields are minimally
coupled to the ordinary derivatives of the background metric. Here
to analyze the concrete value of $\eta/s$ we concentrate on the
action (\ref{dilaton}) without other matter fields coupled. Thus
from the Einstein equation of motion of the action
(\ref{dilaton})~\cite{Guo:2008eq}
\begin{equation}
h(1)=\frac{8}{l^2}e^{\tau\phi_d(1)},
\end{equation}
we can express the ratio of $\eta/s$ as
\begin{equation}\label{why}
\frac{\eta}{s}=\frac{1}{4\pi}\bigg(1-4\lambda
e^{(\tau-\gamma)\phi_d(1)}\Big(1+2\gamma{\phi_d}'(1)\Big)\bigg).
\end{equation}
It is easy to check that when $\phi_d (1)\rightarrow 0$, (\ref{why})
reduces to $(1-4\lambda)/4\pi$ for the pure Gauss-Bonnet case.
However, it is worth noting that the pure Gauss-Bonnet black hole
solution without the dilaton is not a solution of equations of
motion for the action (\ref{dilaton}) by simply requiring the
dilaton being a constant, because the constant dilaton field
$\phi_d$ does not satisfy its Klein-Gordon equation with the pure
Gauss-Bonnet black hole metric (in fact, in order to satisfy the
Klein-Gordon equation with a constant dilaton, one has to impose the
additional condition $\gamma=0$). In addition, let us mention here
that the ratio (\ref{why}) looks dependent on the dilaton and its
derivative on the horizon. In fact, the dependence of the derivative
of the dilaton field can be eliminated by its equation of motion.
The Klein-Gordon equation gives~\cite{Guo:2008eq}
\begin{equation}\label{why2}
{\phi_d}'(1)=-12\gamma\lambda
e^{(\tau-\gamma)\phi_d(1)}+\frac{3}{2}\tau,
\end{equation}
on the horizon. After substituting (\ref{why2}) into (\ref{why}), we
get
\begin{equation}\label{why3}
\frac{\eta}{s}=\frac{1}{4\pi}\bigg(1-4\lambda
e^{(\tau-\gamma)\phi_d(1)}\Big(1-24\gamma^2\lambda
e^{(\tau-\gamma)\phi_d(1)}+3\gamma\tau\Big)\bigg).
\end{equation}
This is our main result. Note that (\ref{why3}) is valid for all
Ricci-flat solutions of the action (\ref{dilaton}). When $\phi_d=0$
 and $\gamma=0$, it reduces to the one for pure Gauss-Bonnet
 gravity without dilaton field.

\section{Causal constraints}

It was discovered that the KSS bound can be violated in theories
with higher order gravity corrections \cite{Brigante:2007nu}, but
the causality on the boundary field can give a constraint to the
parameter and thus we can have a new lower bound on
$\eta/s$~\cite{Brigante:2008gz,Brigante:2007nu}. Here we also discuss
the causal problem on the boundary field theory to see what kind of
constraints we can get on the $\eta/s$ ratio.

Following the method used in Ref.~\cite{Brigante:2008gz}, we need to
write down the graviton equation of motion in the form of
\begin{equation}\label{eom_eff}
\bar{g}^{MN}\bar{\nabla}_{M}\bar{\nabla}_{N}\phi=0,
\end{equation}
where $\bar{g}^{\mu\nu}$ is the effective metric defined in
(\ref{gbar}) (different from the $\widetilde{g}_{MN}$ we used above)
describing the motion of graviton, and $\bar{\nabla}$ is the
derivative operator using the metric $\bar{g}^{MN}$. The equation of
motion (\ref{eom_eff}) can be directly derived from the action
(\ref{minimal}) with $\bar{m}=0$. Then we apply the standard
geometrical optics approximation in the large momentum limit. To be
more explicit, we write the wave function in the form
$\phi=\phi_{en}(t,u,z)e^{i\theta(t,u,z)}$, where $\phi_{en}$ denotes
a slowly changing envelope function and $\theta$ is a rapidly
varying phase function. Inserting this into (\ref{eom_eff}), we
obtain at leading order
\begin{equation} \label{tan}
\frac{dx^{M}}{ds}\frac{dx^{N}}{ds}\bar{g}_{MN}=0,
\end{equation}
with the identification $\frac{dx^{M}}{ds}\equiv
\bar{g}^{MN}k_{N}\equiv \bar{g}^{MN}\bar{\nabla}_{N}\theta$.

This equation describes a classical particle moving in the spacetime
with a metric $\bar{g}_{MN}$, which is no longer the same as the
background metric. In the new spacetime $\bar{g}_{MN}$, there are
still translation symmetries in the $t$ and $z$ directions, so
$\omega=i\bar{\nabla}_t\theta$ and $q=-i\bar{\nabla}_{z}\theta$ are
conserved integrals of motion along the geodesic. Then
Eq.~(\ref{tan}) can be expanded as
$$
\bar{g}^{tt}\omega^2+\bar{g}^{zz}q^2+\bar{g}_{uu}(\frac{du}{ds})^2=0,
$$
which can be rewritten as
\begin{equation}
(\frac{du}{ds})^2=(-\bar{g}^{tt}\bar{g}^{uu}q^2)[\frac{\omega^2}{q^2}
-\frac{\bar{g}^{zz}}{-\bar{g}^{tt}}
].
\end{equation}
In the following discussion we assume $q^2>0$ and the term
$(-\bar{g}^{tt}\bar{g}^{uu}q^2)$ is always larger than zero, so we
can rescale $s$ as $\tilde{s}=s\sqrt{-\bar{g}^{tt}\bar{g}^{uu}q^2}$
to absorb this term and get
\begin{equation}
(\frac{du}{d\tilde{s}})^2=\frac{\omega^2}{q^2}-\frac{\bar{g}^{zz}}{-\bar{g}^{tt}}
.
\end{equation}
This equation describes a one-dimensional system with a particle of
energy $\frac{\omega^2}{q^2}$ moving in a potential given by
$\frac{\bar{g}^{zz}}{-\bar{g}^{tt}}$. This will correspond to a
bouncing geodesic starting and ending at the boundary
\cite{Brigante:2008gz}. Note that unlike the case in Ref.~[33],
although $\tilde{s}$ is not an affine parameter, we can still get
the bouncing geodesics.

Our next task is the same as in the case without
the dilaton~\cite{Brigante:2008gz}. Along a bouncing geodesic, we have
\begin{equation}
\triangle
t(\alpha)=2\int^{u_{turn}(\alpha)}_0\frac{\dot{t}}{\dot{u}}du=2\int^{u_{turn}(\alpha)}_0
\sqrt{\frac{-\bar{g}^{tt}}{\bar{g}^{uu}}}\frac{\alpha}{\sqrt{\alpha^2-c_g^2
}}du,
\end{equation}

\begin{equation}
\triangle
z(\alpha)=2\int^{u_{turn}(\alpha)}_0\frac{\dot{z}}{\dot{u}}du=2\int^{u_{turn}(\alpha)}_0
\sqrt{\frac{-\bar{g}^{tt}}{\bar{g}^{uu}}}\frac{c_g^2}{\sqrt{\alpha^2-c_g^2
}}du,
\end{equation}
where
$$\alpha=\frac{\omega}{q}, \ \ \
c_g^2=\frac{\bar{g}^{zz}}{-\bar{g}^{tt}},$$ and dot denotes the
derivative with respect to $s$.

The graviton moving along the bouncing geodesic will hover near
$u_{turn}$ if $\alpha\rightarrow c_{g,max}$. So if $c_g$ has a
maximal value which is larger than $1$ in the bulk region $0<u<1$,
the ratio $\frac{\triangle z}{\triangle t}$ can be larger than $1$.
This means causality violation of the boundary field theory, as
$\frac{\triangle z}{\triangle t}$ describes the effective velocity
of the graviton moving from one point on the boundary, along the
bouncing geodesic, to another point on the boundary. Because $c_g$
is zero on the horizon and is set to $1$ on the boundary, we can
have a peak of $c_g$ which is larger than $1$ if
\begin{equation}
\frac{\partial c_g^2}{\partial
u}\Big|_{u\rightarrow0}=\lim_{u\rightarrow0}\frac{\partial
[\bar{g}^{zz}/(-\bar{g}^{tt})]}{\partial
u}=\lim_{u\rightarrow0}\frac{\partial[\widetilde{g}^{zz}/(-\widetilde{g}^{tt})]}{\partial
u}>0.
\end{equation}
This condition is sufficient but not necessary  for causality
violation.

Let us now calculate $c_g^2$ at $u\rightarrow0$ to give the
constraint. According to Ref.~\cite{Guo:2008eq}, the spacetime is
asymptotically AdS, and we have
\begin{eqnarray}
\label{metrictt}
g_{tt}\Big|_{u\rightarrow0}= - \frac{a_1}{u} + a_2 u + \cdots,\\
\label{metricuu}
g^{uu}\Big|_{u\rightarrow0}= b_1 u^2 + b_2 u^4 + \cdots,\\
\phi_d \Big|_{u\rightarrow0}= \phi_0 + \phi_1 u^{2+\epsilon} +
\cdots,
\end{eqnarray}
where $\epsilon$ is a positive constant depending on some parameters
in this theory, and $\phi_0$ and $\phi_1$ are two constants,
assuming that a nonrational term does not contribute. By using the
Einstein equations and Klein-Gordon equation, we can fix those
expansion parameters as
\begin{eqnarray}
&& a_1 = \frac{r_+^2}{l^2_{\rm eff}}N^2, ~~
a_2 = \frac{2M}{r_+^2}N^2, \nonumber\\
&& b_1 = \frac{4}{l^2_{\rm eff}}, ~~
b_2 = -\frac{8M}{r_+^4},
\label{coe}
\end{eqnarray}
where $l^{-2}_{\rm eff}=\frac{1-\sqrt{1-4\lambda
e^{(\tau-\gamma)\phi_0}}}{2\lambda l^2e^{-\gamma\phi_0}}$,  where
$M$ is the mass parameter of the black hole solution and $r_+$ is
the horizon radius. Here $N^2= l^2_{\rm eff}/l^2$ is introduced in
order to make the bulk metric conformal to a Minkowski spacetime on
the boundary. We can use these asymptotic expansions as well as the
formulas (\ref{tt}) and (\ref{zz}) to expand $c_g^2$ to order of
$u^2$ near the boundary
\begin{eqnarray}
c_g^2\Big|_{u\rightarrow0} &=& \frac{g^{zz}}{-g^{tt}}
\cdot\frac{1-\frac{\lambda l^2}{2}e^{-\gamma\phi_0}(b_1-b_2u^2)
+4b_1\frac{a_2}{a_1}\frac{\lambda l^2}{2}e^{-\gamma\phi_0}u^2
+O(u^{2+\epsilon})} {1-\frac{\lambda l^2}{2}e^{-\gamma\phi_0}(b_1-b_2u^2)
+O(u^{2+\epsilon})} \nonumber \\
&=& \Big(1-\frac{a_2}{a_1}u^2+O(u^3)\Big) \Big(1+\frac{4b_1 \frac{a_2}{a_1}
\frac{\lambda l^2}{2}e^{-\gamma\phi_0}}{1-\frac{\lambda l^2}{2}
e^{-\gamma\phi_0}b_1}u^2+O(u^3)\Big) \nonumber \\
&=& 1+\frac{a_2}{a_1}\left(
\frac{2b_1\lambda l^2e^{-\gamma\phi_0}}{1-\frac{1}{2}b_1\lambda l^2
e^{-\gamma\phi_0}}-1 \right)u^2+O(u^3).
\end{eqnarray}
The terms with $\phi_d'$ or $\phi_d''$ are $O(u^{2+\epsilon})$
terms. Then in order to get $\frac{\partial c_g^2}{\partial u}>0$
near the boundary, we must have
\begin{equation}
\frac{a_2}{a_1}\Big[\frac{2b_1\lambda l^2e^{-\gamma\phi_0}}
{1-\frac{1}{2}b_1\lambda l^2e^{-\gamma\phi_0}}-1\Big]>0.
\end{equation}
Substituting $a_1$, $a_2$ and $b_1$, we get the condition for causal
violation
\begin{equation}
\lambda e^{(\tau-\gamma)\phi_0}>0.09.
\end{equation}
Thus in order not to have causal violation, we have to impose the
condition that $\lambda e^{(\tau-\gamma)\phi_0}<0.09$. This is
almost the same as the constraint for Gauss-Bonnet gravity without
nontrivial dilaton (in which $\lambda<0.09$), except for a shift to
$\lambda$ brought by the dilaton field on the boundary. This is
because the Gauss-Bonnet black hole solution with nontrivial dilaton
has the same asymptotic behavior as the case without the dilaton field.

With this causal constraint, we can come back to analyze the
result~(\ref{why3}). We rewrite (\ref{why3}) as
\begin{equation}
\frac{\eta}{s}=\frac{1}{4\pi}\bigg(1-4\lambda
e^{(\tau-\gamma)\phi_0}
e^{(\tau-\gamma)[\phi_d(1)-\phi_0]}\Big(1-24\gamma^2\lambda
e^{(\tau-\gamma)\phi_d(1)}+3\gamma\tau\Big)\bigg),
\end{equation}
where the constraint is $\lambda e^{(\tau-\gamma)\phi_0}<0.09$.
Remember that to make sure the gravity regime is valid, we should
assume that $e^{\tau\phi_d}$ and $e^{-\gamma\phi_d}$ are of the
order $O(1)$ in the bulk and $\lambda\ll1$. Also we have to assume
that $e^{\phi_0}\ll 1$ in order to make sure the string coupling
constant $\ll 1$, and luckily this can be checked to be the case
after imposing the Klein-Gordon equation of the dilaton field at the
boundary. The term $-24\gamma^2\lambda e^{(\tau-\gamma)\phi_d(1)}$
can be neglected compared to $3\gamma\tau$. Thus it is easy to see
that the new lower bound imposed in Ref.~\cite{Brigante:2007nu} can
be absolutely violated for solutions with
$(\tau-\gamma)[\phi_d(1)-\phi_0]>0$ and $3\gamma\tau>0$. Because we
must have $\tau-\gamma<0$ in order for the black hole solutions to
exist \cite{Guo:2008eq}, the new lower bound imposed in
Ref.~\cite{Brigante:2007nu} can be violated for solutions with
$\phi_d(1)-\phi_0<0$ and $3\gamma\tau>0$. This kind of solutions
indeed exists, as shown in Ref.~\cite{Guo:2008eq}. Both
$e^{(\tau-\gamma)[\phi_d(1)-\phi_0]}$ and $3\gamma\tau$ are of the
order $O(1)$, so the violation is small, of the order $O(\lambda)$.

\section{Conclusions and Discussions}

In this paper we have calculated the shear viscosity of field
theories with gravity duals of Gauss-Bonnet gravity with a
nontrivial dilaton using AdS/CFT and found that the ratio of the
shear viscosity over entropy density explicitly depends on the
dilaton field on the black hole horizon. Also we have discussed the
causal violation condition of the dual field theory and found that
it is the same as the case without the dilaton field in the sense of
rescaling the Gauss-Bonnet coupling and effective cosmological
constant by the dilaton field at the boundary. After imposing causal
constraint for the boundary field theory, we have found that the new
lower bound $4/25\pi$ may have a small violation due to the
nontrivial dilaton.

In a recent paper \cite{Buchel:2008vz}, it was argued that the KSS
bound would be violated for super conformal field theories with
non-equal central charges. They also showed that the scalars and
vectors coupled to the Gauss-Bonnet gravity only affect the value of
the shear viscosity to the order $O(\lambda^2)$. However, in our
case we find that the nontrivial coupled dilaton field affects the
value of the shear viscosity at the order $O(\lambda)$. They are not
inconsistent with each other. In Ref.~\cite{Buchel:2008vz}, the
authors considered perturbative solutions due to scalar fields and
higher derivative curvature terms, based on a five-dimensional AdS
black hole solution. In that case, the scalars should acquire an
expectation value of order $O(\lambda)$ in the AdS vacuum and black
hole backgrounds, so the scalars only affect the value of shear
viscosity at the order of $O(\lambda^2)$.  In our case, we
considered exact solutions of equations of motion. The scalar field
is of the order $O(1)$, and the shear viscosity is determined by the
value of the scalar on the horizon.  The violation to the new lower
bound can be of order $O(\lambda)$.  Note that the dual field theory
will be a conformal field theory only at the UV boundary. Therefore
it is expected that the bulk viscosity for the dual field theory of
this black hole solution is also nonzero. It would be interesting to
calculate the bulk viscosity in this background using the sound mode
to see if the bulk viscosity has any nontrivial correction from the
nontrivial dilaton coupling.

\section*{Acknowledgements}
We thank Dr. Z. K. Guo for useful discussions. This work was
supported in part by a grant from the Chinese Academy of Sciences
with No. KJCX3-SYW-N2,  grants from NSFC with No. 10821504 and No.
10525060. The work of N.O. was supported in part by the Grant-in-Aid
for Scientific Research Fund of the JSPS No. 20540283 and by the
Japan-U.K. Research Cooperative Program.

\end{document}